\documentclass[lettersize,journal]{IEEEtran}

\usepackage{amsmath,amsfonts}
\usepackage{algorithmic}
\usepackage{algorithm}
\usepackage{array}
\usepackage[caption=false,font=normalsize,labelfont=sf,textfont=sf]{}
\usepackage{textcomp}
\usepackage{stfloats}
\usepackage{color}
\usepackage{lipsum}  
\usepackage{url}
\usepackage{verbatim}
\usepackage{graphicx}
\usepackage{mhchem}
\usepackage{cite}
\usepackage[export]{adjustbox}
\usepackage[font=small,labelfont=bf]{caption}
\usepackage{subcaption}
\usepackage{layouts}
\usepackage{blindtext}
\usepackage{multirow}
\usepackage{wrapfig}
\DeclareMathOperator{\sinc}{sinc}
\DeclareMathOperator{\erf}{erf}
\usepackage{float}

\begin{document}

\title{Frequency-Domain Model of Microfluidic Molecular Communication Channels with Graphene BioFET-based Receivers}
\author{Ali Abdali,~\IEEEmembership{Student Member,~IEEE},
        and Murat Kuscu,~\IEEEmembership{Member,~IEEE}
       \thanks{The authors are with the Nano/Bio/Physical Information and Communications Laboratory (CALICO Lab), Department of Electrical and Electronics Engineering, Koç University, Istanbul, Turkey (e-mail: \{aabdali21, mkuscu\}@ku.edu.tr).}
	   \thanks{This work was supported in part by EU Horizon 2020 MSCA-IF under Grant \#101028935, and by The Scientific and Technological Research Council of Turkey (TUBITAK) under Grant \#120E301.}}



\maketitle

\begin{abstract}
Molecular Communication (MC) is a bio-inspired communication paradigm utilizing molecules for information transfer. Research on this unconventional communication technique has recently started to transition from theoretical investigations to practical testbed implementations, primarily harnessing microfluidics and sensor technologies. Developing accurate models for input-output relationships on these platforms, which mirror real-world scenarios, is crucial for assessing modulation and detection techniques, devising optimized MC methods, and understanding the impact of physical parameters on performance. In this study, we consider a practical microfluidic MC system equipped with a graphene field effect transistor biosensor (bioFET)-based MC receiver as the model system, and develop an analytical end-to-end frequency-domain model. The model provides practical insights into the dispersion and distortion of received signals, thus potentially informing the design of new frequency-domain MC techniques, such as modulation and detection methods. The accuracy of the developed model is verified through particle-based spatial stochastic simulations of pulse transmission in microfluidic channels and ligand-receptor binding reactions on the receiver surface. 
\end{abstract}


\section{Introduction}

\IEEEPARstart {M}{olecular} Communications (MC) is a bio-inspired communication paradigm that uses molecules as information carriers \cite{akan2016fundamentals}. The unique properties of MC, such as biocompatibility, energy efficiency, and reliability under complex and dynamic physiological conditions, are promising for enabling seamless interactions among natural/synthetic cells and micro/nanoscale devices, so-called bio-nano things. Through the emerging Internet of Bio-Nano Things (IoBNT) framework, MC is expected to usher in a new era of unparalleled healthcare and environmental applications at the intersection of information communication technologies, biotechnology, and nanotechnology \cite{akyildiz2015internet,akyildiz2020panacea}. 

MC research has predominantly focused on the development of theoretical channel models, modulation, detection and coding schemes, as well as the design of transmitter and receiver architectures \cite{kuscu2019transmitter,jamali2019channel}. Recent progress in the field has facilitated the integration of experimental validations with theoretical studies, utilizing MC testbeds of varying scales and sophistication. Notably, some of these testbeds, due to their scalability to micro/nanoscales, have the potential to serve as an ideal link between theoretical frameworks and practical applications of MC. Microfluidics technology plays a pivotal role in these practical investigations, as it enables the testing of diverse MC channels while offering comprehensive control over system parameters, such as flow conditions and channel geometry. Moreover, microfluidic channels closely mimic blood vessels and other biological microenvironments, characterized by convection-diffusion-based molecular transport processes \cite{zadeh2023microfluidic}. 

Integrating chemical sensors into microfluidic chips has augmented the utility of these testbeds, with sensors acting as MC receivers that vary in material, geometry, and transduction processes. Among these, affinity-based field-effect transistor biosensors (bioFETs) have emerged as compelling MC receiver architectures due to their inherent signal amplification, miniaturization capabilities, and ligand receptor-based interfaces that provide control over selectivity and sensitivity, resembling biological cells performing molecular sensing. Graphene bioFETs have garnered particular attention owing to the graphene's flexibility, two-dimensional (2D) geometry, and capacity to be functionalized with various bioreceptors, including DNA aptamers and proteins \cite{civas2023graphene}. Initial investigations involving practical microfluidic MC systems equipped with graphene bioFET-based MC receivers have already unveiled crucial insights into the effects of convection, diffusion, ligand-receptor (LR) binding reactions, and receiver material properties on the MC performance \cite{kuscu2021fabrication}. 

Despite these developments, the majority of theoretical and practical studies still primarily focus on the time-domain aspects of MC systems. This can be attributed to the fundamentally distinct nature of the information carriers, i.e., discrete molecules, which lead to ambiguities and complications in defining carrier waves and frequencies for this unconventional communication technique. Additionally, the innate nonlinearity and time-variance of MC communication systems pose challenges to the adoption of frequency-domain techniques as widely-utilized tools in MC research. Nevertheless, when operating regimes can be characterized or approximated as linear and time-invariant (LTI), exploring the frequency-domain features of MC systems can yield crucial insights regarding channel characteristics, such as bandwidth, as well as dispersion and distortion of the transmitted signals. This approach also offers a deeper understanding of the impact of various system parameters on communication performance, including channel geometry, LR binding kinetics at the channel/receiver interface, and the electrical characteristics of the transducer channel within the receiver. Moreover, frequency-domain models can enable the adoption of sophisticated communication tools and methods from conventional EM, including transfer functions and filters, to optimize MC systems and develop new communication techniques, such as frequency-domain pulse-shaping, modulation and detection techniques. 

There has been a limited focus on frequency-domain analysis in MC. A notable contribution is the frequency-domain model for diffusion-based MC systems developed in \cite{pierobon2010physical}, which allows the determination of the end-to-end normalized gain and delay of the MC system as a function of frequency. In \cite{huang2021frequency}, a frequency-domain equalizer (FDE) was proposed to address the inter-symbol interference (ISI) problem in MC. The transfer function of the MC channel, considering only diffusion-based transport, was derived in \cite{wang2014transmit}. 
In our recent study \cite{civas2023frequency}, we introduced a frequency-domain detection technique for MC to estimate the concentration of information molecules in the presence of interfering molecules, leveraging LR binding kinetics. This method employs the power spectral density (PSD) of binding noise, which exhibits unique properties for each molecule type, enabling the differentiation of information molecules from interferers in the frequency-domain.

In this study, we present an end-to-end frequency-domain system model for a microfluidic MC channel employing a graphene bioFET-based MC receiver with ligand receptors on its surface for detecting molecular messages carried by information molecules (i.e., ligands). We consider transmitted signals as finite-duration molecular concentration pulses. We partition the end-to-end MC system into three subsystems: (i) the microfluidic propagation channel, where ligand propagation is governed by convection and diffusion; (ii) the channel/receiver interface, where the receiver's surface receptors interact with propagating ligands; and (iii) the graphene bioFET-based receiver, which transduces the number of bound receptors into an output electrical current. By employing LTI approximations, we analyze each subsystem independently and derive their transfer functions. We then combine these to obtain the end-to-end MC system's transfer function. The developed frequency-domain model is validated through particle-based spatial stochastic simulations using \emph{Smoldyn}, an open-source simulation framework \cite{smoldyn2022}. The simulation results show a strong agreement with the developed analytical frequency-domain model. We also examine the impact of various system parameters, such as pulse width of input signals, diffusion coefficient of ligands, and binding and unbinding rates of LR pairs, on the transfer function. Additionally, we leverage the developed model to determine the minimum sampling frequency for digitizing the output current by identifying the cutoff frequency and applying the Nyquist–Shannon theorem.

The remainder of this paper is organized as follows. Section \ref{endtoend} offers an in-depth analysis of the three key components of the microfluidic MC system, followed by the development of the end-to-end frequency-domain model, which is then utilized to obtain the output signal. Section \ref{simulation} presents the simulation results intended to validate the developed model. Lastly, Section \ref{conclusion} delivers concluding remarks.

\section{End-to-End Frequency-Domain Model}
\label{endtoend}
\begin{figure*}[t]
    \centering
    \includegraphics[scale=0.65]{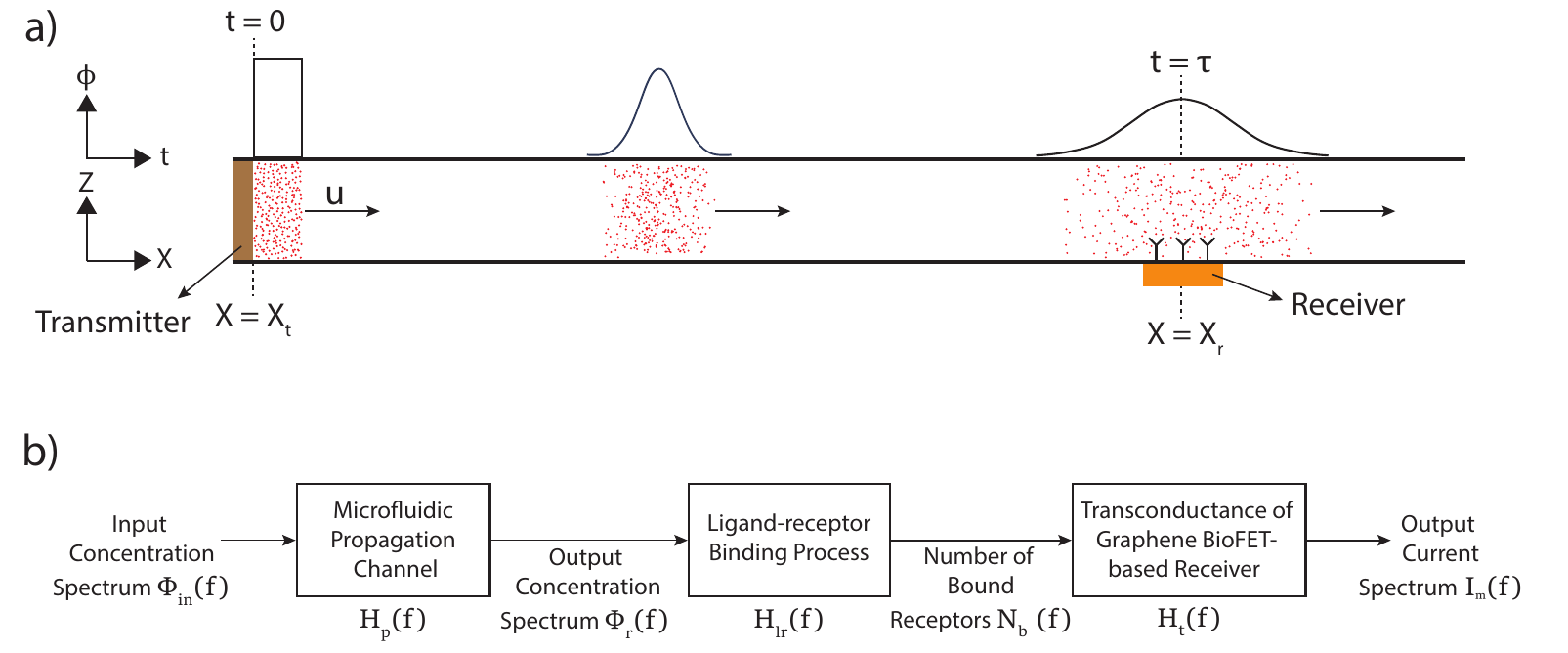}
    \caption{(a) 2D view of microfluidic propagation channel. Propagation of concentration signal from the transmitter, positioned at $x=x_\mathrm{t}$, to the receiver, positioned at $x=x_\mathrm{r}$, based on convection and diffusion is depicted. (b) Block diagram of microfluidic MC channel utilizing LR interactions in biorecognition layer of graphene bioFET-based MC receiver. }
    \label{block_diag}
\end{figure*}

In this section, we present the derivation of the end-to-end frequency-domain model for the microfluidic MC system, depicted in Fig. \ref{block_diag}(a). The system comprises a rectangular cross-section microfluidic channel in which molecular signals are uniformly transmitted across the cross-section of the channel inlet. The microfluidic channel is assumed to be open-ended, with a two-dimensional graphene bioFET-based biosensor serving as the receiver, positioned at the bottom of the channel without obstructing molecular propagation. 

To establish the end-to-end model, transfer functions are derived for three subsystems:  propagation of ligands within the microfluidic MC channel, LR binding interactions at the receiver surface, and molecular-to-electrical transduction process within the graphene bioFET-based MC receiver. The block diagram illustrating the end-to-end microfluidic MC system is provided in Fig. \ref{block_diag}(b).

\subsection{Transfer Function of Microfluidic Propagation Channel}
We consider a straight microfluidic channel with a rectangular cross-section which is filled up with an electrolyte as the medium of propagation for molecular signals. The transmitter is located at the entrance of the microfluidic channel, while the receiver, a graphene bioFET, is situated at the base of the channel at position $x=x_\mathrm{r}$ as shown in Fig. \ref{block_diag}(a). The surface of the graphene transduction channel of the receiver is functionalized with selective receptors, which are exposed to ligands of time-varying concentration. The receiver senses the concentration of ligands, flowing over its surface, through LR binding reactions. The flow is unidirectional from the inlet to the open outlet of the microfluidic channel. 

The convection-diffusion equation, which is a linear partial differential equation, describes the behavior of mass transport of ligands within the microfluidic channel as \cite{kuscu2018modeling}
\begin{align}
    \frac{\partial\phi}{\partial t}= -u \nabla \phi + D \nabla ^2\phi,
    \label{eq:conv_diff}
\end{align}
where $\phi$ is the concentration of the released ligands, $D$ is the diffusion coefficient of the ligands, $u$ is the fluid flow velocity. The convection-diffusion equation describes the spatiotemporal evolution of the ligand concentration profile, $\phi$, through the convective term $(-u \nabla \phi)$ and the diffusive term $(D \nabla ^2\phi)$. 

In this study, we consider a uniform and unidirectional fluid flow solely along the x-axis, i.e., $u = u_\mathrm{x}$. To simplify the analysis of ligand transport, we adopt a one-dimensional (1D) approximation, focusing primarily on the x-axis. This assumption is justified when the ligand propagation is predominantly unidirectional, and lateral dispersion is much smaller than longitudinal transport. Such conditions arise due to the design and flow characteristics of the microfluidic channel, as discussed in \cite{bicen2013system}. The validity of this approximation can be quantified using the Péclet number, defined as $Pe = u_\mathrm{x} l / D$. In this definition, $l$ denotes the characteristics length, and for our case, corresponds to the distance between the transmitter and the receiver, i.e., $l = x_\mathrm{r}$. When $Pe \gg 1$, the 1D approximation is valid, which is consistent with the model parameter values considered in this study. Accordingly, the 1D solution of \eqref{eq:conv_diff} for an input concentration signal in the form of an impulse at the origin (i.e., $\phi_\mathrm{in}(x, t) = \delta(x - x_0, t - t_0)$ with $x_0 = 0, t_0 = 0$), gives the impulse response of the microfluidic propagation channel:
\begin{align}
    h_\mathrm{p}(x,t)= \frac{1}{\sqrt{4\pi D t}} \exp{\left(-\frac{(x-u_\mathrm{x}t)^2}{4Dt}\right)}.
\end{align}
The propagation delay, $\tau$, is the time it takes for the peak ligand concentration to travel a distance of $x$ from the channel inlet, given by $\tau = \frac{x}{u_\mathrm{x}}$ for $Pe \gg 1$. The received concentration, $\phi_\mathrm{r}$, in the time-domain for an input concentration of $\phi_\mathrm{in}$ in a straight microfluidic MC channel can be calculated through the convolution of the input concentration and the impulse response as follows:
\begin{align}
    \phi_\mathrm{r} (l)=(h\ast \phi_\mathrm{in})(l)= \int^{+\infty}_{-\infty} h_\mathrm{p}(x)\phi_\mathrm{in}(l-x)dx.
    \label{channel_res_t}
\end{align}
For a rectangular finite-duration
concentration pulse input with a pulse width of $T_\mathrm{p}$ and amplitude of $C_\mathrm{m}$, the received concentration at $x=l$ can be calculated using \eqref{channel_res_t} as follows:
\begin{align}
    \phi_\mathrm{r}(l)= \frac{C_\mathrm{m}}{2}\left(\erf{\left(\frac{tu-l+T_\mathrm{p}u}{2\sqrt{Dt}}\right)} -\erf{\left(\frac{tu-l}{2\sqrt{Dt}}\right)}\right).
    \label{t_con_out}
\end{align}
The transfer function of the propagation channel, which represents the frequency response to an impulse signal, can be derived by solving the frequency-domain counterpart of the 1D convection-diffusion equation \eqref{eq:conv_diff} obtained using the Fourier Transform (FT): 
\begin{align}
    j 2\pi f \Phi(x,f) = - u \frac{\partial\Phi(x,f)}{\partial x}+ D \frac{\partial^2\Phi(x,f)}{\partial x^2},
    \label{eq:ft_of_conv_diff}
\end{align}
where, $\Phi(x,f)$, the spectral density of ligand concentration, is obtained by taking the FT of the time-domain ligand concentration signal, i.e., $\Phi(x,f)= \mathcal{F}(\phi(x,t))$ \cite{bicen2013system}. 
Assuming $|(8\pi fD)/u^2|<1$ to have a converging series expansion in solution of \eqref{eq:ft_of_conv_diff}, and fixing $x = x_\mathrm{r}$, the receiver's central position, we can analytically approximate the transfer function of the microfluidic propagation channel, i.e., $H_\mathrm{p}(f)$, as follows \cite{bicen2013system}\\
\begin{align} \label{eq:TF_propagation}
  H_\mathrm{p}(f) &= H_\mathrm{p}(x = x_\mathrm{r},f) \\ \nonumber
 & \approx \exp \biggl(-\left(\frac{(2\pi f)^2D}{u^3}+j\frac{2\pi f}{u}\right) x_\mathrm{r} \biggr).
\end{align}

Spectral density of the received concentration, $\Phi_\mathrm{r}(f) = \Phi(x=x_\mathrm{r}, f)$, can be obtained via multiplication of the transfer function, $H_\mathrm{p}(f)$, and the spectral density of the input ligand concentration signal, $\Phi_\mathrm{in}(f) = \Phi_\mathrm{in}(x = 0, f)$, as follows 
\begin{align}
    \Phi_\mathrm{r}(f) \approx H_\mathrm{p}(f)\Phi_\mathrm{in}(f).
    \label{con_out}
\end{align}
Note that the spectral density of a rectangular finite-duration concentration pulse input signal with amplitude $C_\mathrm{m}$ and pulse width $T_\mathrm{p}$ can be obtained by taking FT as
\begin{align}
    \Phi_\mathrm{in}(f)= \mathcal{F}\big\{C_\mathrm{m} \text{rect}(t/T_\mathrm{p} - 0.5)\big\} = C_\mathrm{m} T_\mathrm{p} \sinc{\left(f T_\mathrm{p} \right)},
    \label{con_in}
\end{align}
where $\text{rect}(t) = 1$ for $-0.5 < t < 0.5$ is the rectangular function. Therefore,  combining \eqref{eq:TF_propagation}, \eqref{con_out}, and \eqref{con_in}, spectral density of the received ligand concentration signal can be approximated as follows:

\begin{align} \label{f_rect}
  \Phi_\mathrm{r}(f) \approx~ &C_\mathrm{m} T_\mathrm{p} \sinc{\left(f T_\mathrm{p}\right)} \\ \nonumber
 & \times \exp \biggl(-\left(\frac{(2\pi f)^2D}{u^3}+j\frac{2\pi f}{u}\right) x_\mathrm{r} \biggr).
\end{align}
\subsection{Transfer Function of the Ligand-Receptor Binding Process}

Propagating ligands encoding information bind to the receptors on the graphene bioFET-based MC receiver surface randomly and reversibly such that a formed LR complex dissociates after a random time duration. In the case of a monovalent reaction, where receptors have only one binding site and can be in one of the two states, unbound (U) or bound (B), the reversible LR binding interactions can be described in terms of reaction rates as follows
\begin{align}
    \ce {U <=>[\ce{\phi_r(t) k_+}][\ce{k_-}] B}, 
\end{align}
where $k_\mathrm{+}$ and $k_\mathrm{-}$ are the binding and unbinding rates of the LR pair, and $\phi_\mathrm{r}(t)$ is the time-varying ligand concentration in the vicinity of receptors \cite{kuscu2019channel}, assuming that receptors are exposed to the same ligand concentration at all times, and the number of ligands bound to receptors is much lower than the number of ligands in the vicinity of the receptors such that the ligand concentration $\phi_\mathrm{r}(t)$ can be assumed to remain unchanged during the LR binding interactions. The number of bound receptors as a function of time, i.e., $N_\mathrm{b}(t)$, can then be written as 
\begin{equation}
\label{eq:nonlin_lr_binding}
\frac{dN_\mathrm{b}(t)}{dt}= k_\mathrm{+} (N_\mathrm{r} - N_\mathrm{b}(t)) \phi_\mathrm{r}(t) - k_\mathrm{-} N_\mathrm{b}(t),
\end{equation}
where $N_\mathrm{r}$ is the total number of receptors on the receiver surface. The second-order reaction represented by the above nonlinear equation can be simplified as a first-order reaction, if the total number of receptors is much higher than the number of bound receptors at all times \cite{lauffenburger1996receptors}, yielding a linear equation: 
\begin{equation}
\label{eq:lin_lr_binding}
\frac{dN_\mathrm{b}(t)}{dt}= k_\mathrm{+} N_\mathrm{r} \phi_\mathrm{r}(t) - k_\mathrm{-} N_\mathrm{b}(t).
\end{equation}
The condition of first-order reaction can be quantitatively formulated as $k_\mathrm{+}\phi_\mathrm{r}(t)\ll k_\mathrm{-}$ \cite{shahmohammadian2013modelling}, ensuring the number of bound receptors is comparatively low with a high unbinding rate. The transfer function of the LR binding process can then be obtained by solving the frequency-domain equivalent of \eqref{eq:lin_lr_binding}: 
\begin{equation}
\label{eq:FT_lr_binding}
j 2 \pi f N_\mathrm{b}(f) = k_\mathrm{+} N_\mathrm{r} \Phi_\mathrm{r}(f) - k_\mathrm{-} N_\mathrm{b}(f),
\end{equation}
considering $\Phi_\mathrm{r}(f)$ as the input signal and $N_\mathrm{b}(f)$ as the output signal: 
\begin{equation}
   H_{\mathrm{lr}}(f) = \frac{k_\mathrm{+} N_\mathrm{r}}{k_\mathrm{-} + j 2 \pi f}.
    \label{eq:TF_lr_binding}
\end{equation}
The transfer function of the LR binding process corresponds to that of a low-pass filter, characterized by a cutoff frequency of $f_\mathrm{c,lr} = k_\mathrm{-}/2\pi$. 
Utilizing \eqref{eq:TF_lr_binding}, the spectral density of the output signal, i.e., time-varying number of bound receptors, can be obtained as follows:
\begin{equation}
   N_\mathrm{b}(f) = H_\mathrm{lr}(f) \Phi_\mathrm{r}(f) = \frac{k_\mathrm{+} N_\mathrm{r}}{k_\mathrm{-} + j 2 \pi f} \Phi_\mathrm{r}(f).
    \label{eq:number_bound}
\end{equation}

\subsection{Transfer Function of Graphene BioFET-based MC Receiver}

\begin{figure}[t]
    \centering
    \includegraphics[scale=0.52
]{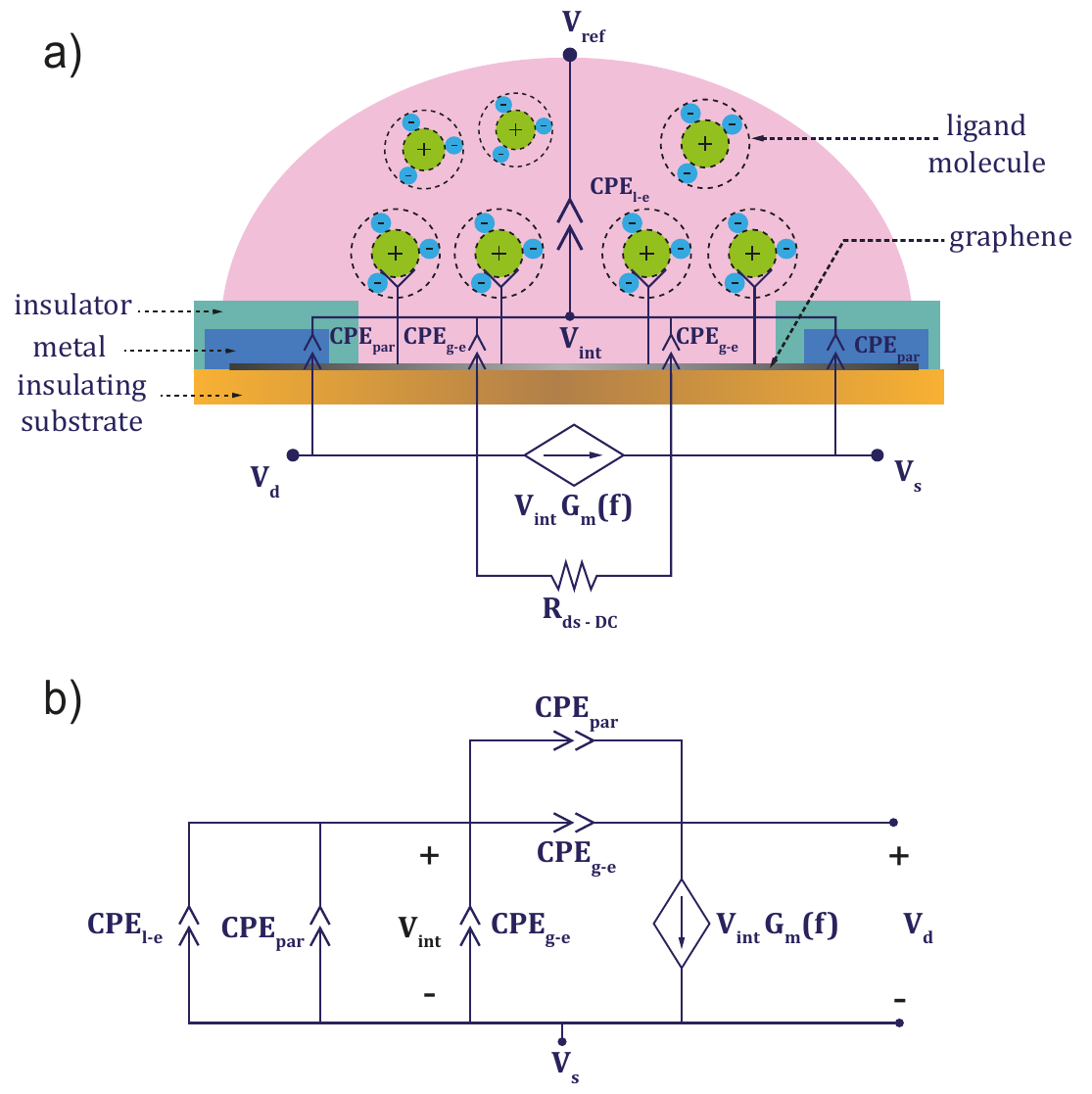}
    \caption{(a) Schematic of the graphene bioFET combined with the equivalent circuit. (b) Small-signal equivalent model of the graphene bioFET.}\vspace{4mm}
\label{biofet}
 \centering
\end{figure}

We consider a graphene bioFET-based MC receiver fabricated on a Si/SiO$_2$ substrate with a monolayer graphene, which is connected to power sources via deposited metal (Cr/Au) drain and source contacts insulated from the electrolyte MC channel through an insulator layer (e.g., thin Al$_2$O$_3$ film), as shown in Fig. \ref{biofet}(a). A bio-recognition layer is incorporated onto the surface of graphene, comprising receptors that interact with ligands through the LR binding process. A DC potential, denoted as $V_\mathrm{ref}$, is applied to the electrolyte to determine the operating point. This receiver architecture has been previously implemented by our group and its fabrication methodology was detailed in \cite{kuscu2021fabrication}. In this architecture, binding of charged ligands to the receptors attached uniformly to the graphene surface results in the modulation of the charge carrier density of the transducer channel, i.e., graphene, through electric field effect. This change in charge carrier density modulates the conductance of the channel, and hence the drain-to-source current ($\Delta I_\mathrm{ds}$) of the receiver. Therefore, the alteration in $\Delta I_\mathrm{ds}$ under constant drain-to-source bias $(V_\mathrm{ds})$ becomes a function of the number of bound ligands (which is equal to the number of bound receptors $N_\mathrm{b}(t)$) and the electrical charge of the bound ligands. Therefore, the bound ligands can be considered functionally equivalent to the gate of the transistor.

In conventional FETs, the effect of gate potential on $\Delta I_\mathrm{ds}$ is quantified through the transconductance of the transistor, denoted by $G_\mathrm{m}$. Likewise, the impact of ligands bound to receptors on $\Delta I_\mathrm{ds}$ can be measured through transconductance. Therefore, transconductance plays a vital role in shaping the input-output relationship of the MC receiver, and the frequency-domain representation of the transconductance, denoted as $G_\mathrm{m}(f)$, becomes a key part of the transfer function of the MC receiver, referred to as $H_\mathrm{t}(f)$. Nevertheless, there is an additional component that contributes to $H_\mathrm{t}(f)$, which will be further explained.

To obtain the $G_\mathrm{m}(f)$, we can use a small-signal model, which is an AC equivalent circuit that approximates the nonlinear behavior of the device with linear elements. In this study, we build on the small-signal model developed in \cite{garcia2020distortion} for graphene solution-gated FETs, used as a neural interface, to obtain the input-output relation in frequency-domain with the input being the time-varying number of bound receptors and the output being the drain-to-source current. Fig. \ref{biofet}(a) presents the schematic of the MC receiver combined with the equivalent circuit to depict physical origin of each element, and Fig. \ref{biofet}(b) demonstrates  the small-signal model of the MC receiver.

We start modeling the MC receiver by investigating solid-liquid interface behavior. The interface of a charged surface and an electrolyte is commonly referred to as an electrical double layer (EDL) \cite{stojek2010electrical}. The electrons on the charged surface and the ions on the electrolyte are separated by a single layer of solvent molecules that stick to the charged surface and act as a dielectric in a conventional capacitor. Hence, EDL properties are generally modeled as a capacitor in the literature \cite{khademi2020structure}. In this model, however, the graphene-electrolyte interface is described as a constant phase element (CPE) (i.e., $CPE_\mathrm{g-e}$) rather than an ideal capacitor to precisely characterize the response of the EDL in graphene bioFETs. The CPE behavior of the graphene-electrolyte interface stems from the presence of charged impurities on the substrate and structural imperfections within the graphene lattice, potentially resulting in a non-uniform local density of states (DOS) \cite{sun2019unique}. 

The impedance of a CPE can be written as follows \cite{barsoukov2005impedance}:
\begin{align}
    Z=\frac{1}{Q_0(j2\pi f)^\alpha},
    \label{Z_CPE}
\end{align}
where $Q_0$ is the admittance at $f=1/2\pi$ Hz and $\alpha$ is a parameter that determines the phase angle. The values of both $Q_0$ and $\alpha$ depend on the applied voltage (which results from the bound charged ligands) and reflect the properties of the graphene-electrolyte interface. A CPE with $\alpha=1$ behaves like an ideal capacitor, while a CPE with $\alpha=0$ behaves like a pure resistor. A CPE with $0<\alpha<1$ represents an imperfect capacitor that has a non-constant capacitance value. The capacitance of a CPE (i.e., $C_\mathrm{CPE}$) can be calculated by equating the imaginary part of the impedance of CPE to the impedance of an ideal capacitor, as proposed by Hsu et al. \cite{hsu2001concerning}. This approach yields the following expression for the capacitance of a CPE: 
\begin{equation}
    C_\mathrm{CPE}=\frac{Q_0}{(2\pi f) ^{1-\alpha}}\exp{\left(j\frac{\pi}{2}(\alpha -1)\right)}.
    \label{CPE}
\end{equation}


The bio-recognition layer can be modeled as a charged capacitor according to Xu et al. \cite{xu2017real}. This represents the double-layer capacitance between a single ligand and electrolyte. In this model, the ligand-electrolyte interface is described as a CPE, which is denoted as $CPE_\mathrm{l-e}$, with $\alpha=1$ to mimic the behavior of an ideal capacitor and ensure notational consistency within the overall model.
When charged ligands bind to receptors, they generate a small signal variation in the gate potential. This variation is transduced into a voltage at the graphene-electrolyte interface ($V_\mathrm{int}$).
A current source element $V_\mathrm{int}G_\mathrm{m}(f)$ is used to model the conversion of AC signals at the gate into AC signals in the drain current ($I_\mathrm{ds}$), where $G_\mathrm{m}(f)$ is the transconductance of the bioFET. To account for the DC current flowing through the graphene bioFET caused by the reference voltage ($V_\mathrm{ref}$), a resistive element $R_\mathrm{ds–DC}$ is employed in the model. However, during small signal analysis, when $V_\mathrm{ref}$ is set to zero, the $R_\mathrm{ds-DC}$ is removed from the small signal model depicted in Fig \ref{biofet}(b). To account for parasitic capacitances in the device that arise as a result of the coupling between electrolyte and the contact metals through the insulating layer, another CPE, $CPE_\mathrm{par}$, is included in parallel with $CPE_\mathrm{g–e}$. As it will be revealed, this CPE affects the high-frequency response of the bioFET. Using this equivalent circuit, we can obtain the frequency-domain representation of transconductance as follows \cite{garcia2020distortion}:
\begin{equation}
    G_\mathrm{m}(f)=\frac{dI_\mathrm{ds}}{dV_\mathrm{int}}|_{v_\mathrm{ds}} + G_\mathrm{m,eff}.
 \label{GM}
\end{equation}

The derivative term on the RHS of \eqref{GM} represents the intrinsic transconductance, which is the change in the drain current with respect to the interface potential. The intrinsic transconductance depends on the interface capacitance between the graphene and the electrolyte ($CPE_\mathrm{g-e}$), which reflects the charge accumulation at the interface. This relationship is given by
\begin{equation}
    \frac{dI_\mathrm{ds}}{dV_\mathrm{int}}|_{v_\mathrm{ds}} = V_\mathrm{ds}\frac{w_\mathrm{g}}{l_\mathrm{g}}\mu_\mathrm{g} C_\mathrm{{CPE_{g-e}}},
    \label{GMINT}
\end{equation}
where $w_\mathrm{g}$ and $l_\mathrm{g}$ represent the width and length of the graphene transduction channel, respectively, and $\mu_\mathrm{g}$ denotes the charge carrier mobility of graphene \cite{xu2011top}. The interface capacitance (i.e., $C_\mathrm{{CPE_{g-e}}}$) can be obtained using \eqref{CPE} to have a  frequency-dependent relation for the intrinsic transconductance.

\begin{figure}[t!]
    \centering
    \includegraphics[scale=0.5]{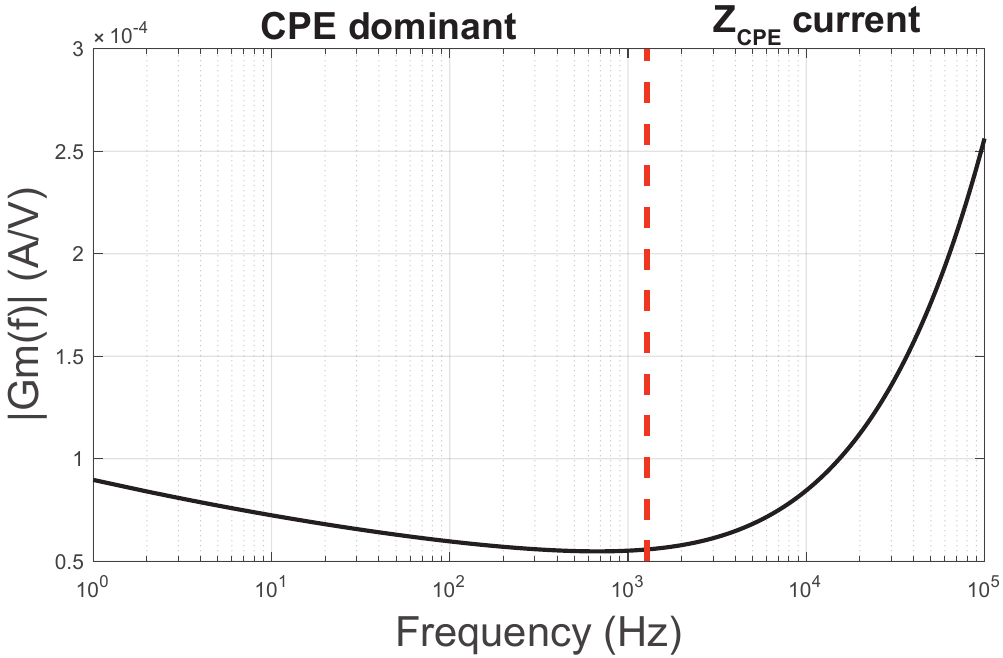}

    \caption{Transconductance $(G_\mathrm{m}(f))$ of graphene bioFET composed of two different response regime: (i) CPE dominant regime and (ii) Z$_\mathrm{CPE}$ current regime.}
    \label{graph}
\end{figure}

An additional term, $G_\mathrm{{m,eff}}$, contributes positively to the gain of the transduction process at high frequencies. The interface capacitances $C_\mathrm{{CPE_{g-e}}}$ and $C_\mathrm{{CPE_{par}}}$ lead to a direct capacitive current between the gate and the graphene bioFET contacts, which is evenly distributed to the drain and source. This contribution, independent of field-effect coupling, can be regarded as an effective transconductance term. As shown in Fig. \ref{graph},  which plots the magnitude of $\lvert G_\mathrm{m}(f) \rvert$ over a range of frequencies for MC receiver, the capacitive contribution to the drain current dominates the frequency response beyond a certain frequency threshold. This contribution can be expressed as \cite{garcia2020distortion}:
\begin{equation}
    G_\mathrm{m,eff}(f)=1/(2Z_\mathrm{{CPE_{g-e}}})+1/(2Z_\mathrm{{CPE_{par}}}).
    \label{Gmeff}
\end{equation}
By explicitly incorporating the frequency dependence in \eqref{Z_CPE} for \eqref{Gmeff}, the second term in \eqref{GM} can be derived as follows:
\begin{align} \label{Gmeff_f}
    G_\mathrm{m,eff}(f) =& Q_\mathrm{g-e}+(2\pi f)^{\alpha _\mathrm{g-e}}e^{j\frac{\pi}{2}\alpha _\mathrm{g-e}} \\ \nonumber
    &+Q_\mathrm{par}(2\pi f)^{\alpha _\mathrm{par}}e^{j\frac{\pi}{2}\alpha _\mathrm{par}}.
\end{align}\\
By combining \eqref{GM} with \eqref{GMINT}, and \eqref{Gmeff_f}, the frequency-dependent transconductance of a MC receiver can be expressed as:
\begin{align}
    &G_\mathrm{m}(f) = \pm V_\mathrm{ds}\frac{w_\mathrm{g}}{l_\mathrm{g}}\mu_\mathrm{g}\frac{Q_\mathrm{g-e}}{(2\pi f)^{1-\alpha_\mathrm{g-e}}}e^{j\frac{\pi}{2}(\alpha_\mathrm{g-e}-1)}\\ \nonumber
    &+Q_\mathrm{g-e}(2\pi f)^{\alpha _\mathrm{g-e}}e^{j\frac{\pi}{2}\alpha _\mathrm{g-e}}+Q_\mathrm{par}(2\pi f)^{\alpha _\mathrm{par}}e^{j\frac{\pi}{2}\alpha _\mathrm{par}}.
\end{align}

The equation above uses the $\pm$ sign, which is positive in the electron conduction regime, and negative in the hole conduction regime of the bioFET. In this study, we focused on the hole conduction regime when plotting $\lvert G_\mathrm{m}(f) \rvert$ and conducting the simulations. In Fig. \ref{graph}, two distinct response regimes can be identified: (i) a \emph{CPE dominant} regime (up to 1 kHz), and (ii) a \emph{Z$_\mathrm{CPE}$ current} regime where $G_\mathrm{m}$ increases due to capacitive currents (above 1 kHz).

To obtain the transfer function of the bioFET-based MC receiver, i.e., $H_\mathrm{t}(f)$, in addition to $G_\mathrm{m}(f)$, we need to derive the potential created by a single ligand on the graphene surface ($V_\mathrm{int}$). As it will be revealed, by including $V_\mathrm{int}(f)$ in the $H_\mathrm{t}(f)$, we will be able to derive the spectral density of the output current, i.e., $I_\mathrm{m}(f)$, through end-to-end transfer function. The effective charge on the graphene surface resulting from binding of each ligand to the receptor is determined by the expression $Q_\mathrm{m}=q_\mathrm{eff}N_\mathrm{e^{-}}$, where $N_{e^{-}}$ denotes the average number of free electrons per ligand. The mean effective charge, $q_\mathrm{eff}$, represents the charge that a single electron of a ligand can generate on the graphene surface in the presence of ionic screening in the medium. The relationship is given by $q_\mathrm{eff}= q \times \exp{(-r/\lambda_\mathrm{D})}$ where $q$ is the elementary charge and $r$ represents the average distance between the ligand electrons in the bound state and the surface of the transducer. It is assumed that this average distance is equivalent to the average length of the surface receptor in the bound state \cite{kuscu2016modeling}. The Debye length, $\lambda_\mathrm{D}$, characterizes the ionic strength of the medium, and is given by $\lambda_\mathrm{D}=\sqrt{(\epsilon_\mathrm{M}k_\mathrm{B}T)/(2 N_\mathrm{A}q^2c_\mathrm{ion})}$, where $\epsilon_\mathrm{M}$ is the dielectric permittivity of the medium, $k_\mathrm{B}$ is Boltzmann’s constant, $T$ is the temperature, $N_\mathrm{A}$ is Avogadro’s number, and $c_\mathrm{ion}$ is the ionic concentration of the medium \cite{rajan2013performance}. Finally, the interface potential generated by the charge accumulated on the surface by a single ligand is as follows \cite{kuscu2016modeling}:
\begin{equation}
    V_\mathrm{int}(f)= \frac{Q_\mathrm{m}}{C_\mathrm{CPE_{eq}}},
    \label{vint}
\end{equation}
where $C_\mathrm{CPE_{eq}}$ is the equivalent capacitance of the transducer, which is comprised of a parallel combination of $CPE_\mathrm{l-e}$, $CPE_\mathrm{g-e}$, and $CPE_\mathrm{par}$ connected in series with another parallel pair of $CPE_\mathrm{g-e}$ and $CPE_\mathrm{par}$ as shown in Fig. \ref{biofet}(b). This can be expressed as:
\begin{align}
    C_\mathrm{CPE_{eq}}=& \left(\frac{1}{C_\mathrm{CPE_{l-e}}+C_\mathrm{CPE_{g-e}}+C_\mathrm{CPE_{par}}}\right.\\ \nonumber
    +&\left. \frac{1}{C_\mathrm{CPE_{g-e}}+C_\mathrm{CPE_{par}}}\right)^{-1}, 
\end{align}
where frequency-dependent relation of all $C_\mathrm{CPE}$ terms can be obtained by utilizing \eqref{CPE}. Therefore, the transfer function of the transduction process in a graphene bioFET-based MC receiver, i.e., $H_\mathrm{t}(f)$, can be written by using \eqref{GM} and \eqref{vint} as
\begin{align}
    H_\mathrm{t}(f) = V_\mathrm{int}(f) G_\mathrm{m}(f). 
    \label{H3}
\end{align}

\subsection{End-to-End Transfer Function and Output Current Spectral density}
The end-to-end transfer function of a microfluidic MC channel with graphene bioFET-based receiver can be expressed using Equations \eqref{eq:TF_propagation}, \eqref{eq:TF_lr_binding}, and \eqref{H3} as follows
\begin{align}
    H(f) &= H_\mathrm{p}(f) \times H_\mathrm{lr}(f) \times H_\mathrm{t}(f) \\ \nonumber
    & =  V_\mathrm{int}(f) G_\mathrm{m}(f) \left(\frac{k_\mathrm{+} N_\mathrm{r}}{k_\mathrm{-} + j 2 \pi f}\right) e^{-\left(\frac{(2\pi f)^2D}{u^3}+j\frac{2\pi f}{u}\right)x_\mathrm{r}}.
\end{align}
Spectral density of the output current can be obtained by using the end-to-end transfer function and the spectral density of the input concentration signal as: 
\begin{equation}
    I_\mathrm{m}(f) = H(f) \times \Phi_\mathrm{in}(f). 
    \label{Im}
\end{equation}

\section{Numerical Results} 
\label{simulation}
In this section, we present the numerical results obtained using the developed analytical frequency-domain model, which is validated through particle-based simulations under various settings. The default values for the parameters used in the analyses are provided in Table \ref{tab:parameters}. The admittance and phase angle values for $CPE_\mathrm{g-e}$ and $CPE_\mathrm{par}$ are extracted from the experimentally fitted data in \cite{garcia2020distortion}, conducted in an electrolyte medium with an ionic strength of 0.5 mM. We considered the same ionic strength ($c_\mathrm{ion}=$ 0.5 mM). As for the admittance of $CPE_\mathrm{l-e}$, it is assigned considering the fact that the area of the double-layer interface between ligands and electrolyte is significantly smaller compared to the double-layer surface at the graphene-electrolyte interface. Consequently, based on the parallel plate capacitor formula $C=\varepsilon \frac{A}{d}$, where $\varepsilon$ is permittivity, and $d$ is the distance between the surface layers (a single layer of molecules in this case), it is evident that the capacitive behavior of $CPE_\mathrm{l-e}$ are significantly lower compared to $CPE_\mathrm{g-e}$ since the values of $A$ and $d$ remain the same for both interfaces. We set $\mu_\mathrm{g}=200$ cm$^2$/Vs as reported in \cite{kuscu2021fabrication}. Aptamers are utilized as the receptors, and their default length is defined as 2 nm \cite{kuscu2016physical}. Binding and unbinding rates, $k_\mathrm{+}$ and $k_\mathrm{-}$ are set considering the assumption of \eqref{eq:lin_lr_binding} and accepted values in the MC literature \cite{pierobon2011noise}. We consider the microfluidic channel with a cross-sectional height of $h_\mathrm{ch}=$ 3 $\mu$m, a width of $w_{ch}=$ 3 $\mu$m, and a length of $l_\mathrm{ch}=$ 200 $\mu$m, resulting in a laminar and steady flow. The simulations were performed using \emph{Smoldyn}, a particle-based spatial stochastic simulation framework that offers high spatiotemporal resolution by simulating each molecule of interest individually \cite{smoldyn2022}. This approach captures the inherent stochasticity of molecular transport and reactions and provides nanometer-scale spatial resolution. 

\begin{figure}[t]
    \centering
    \includegraphics[scale=0.315]{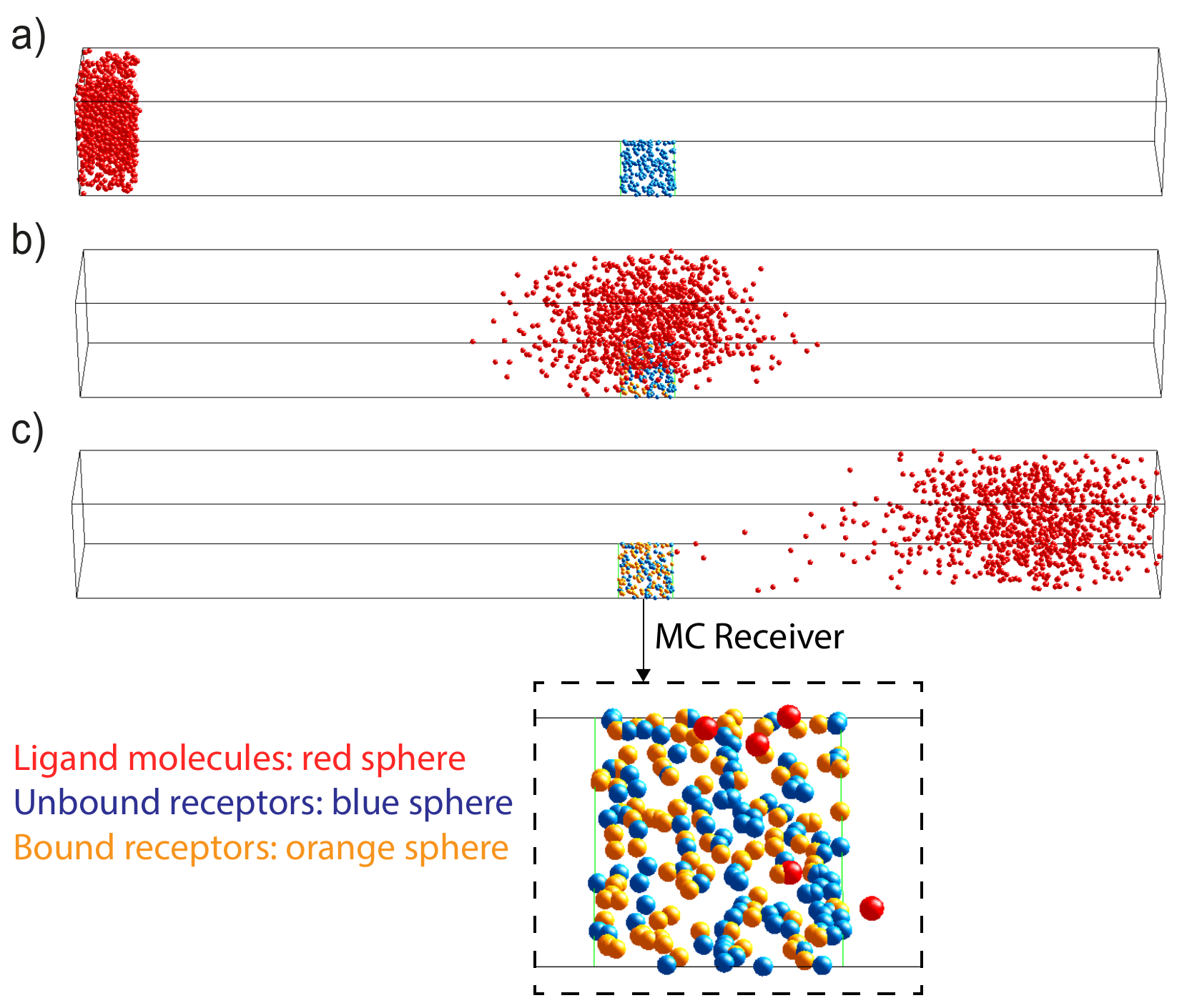}
    \caption{Smoldyn simulation environment: (a) A pulse concentration of ligands is released into the microfluidics MC channel. (b) The pulse around the MC receiver position, where ligands initiate interactions with receptors. The dispersion of the pulse signal occurs as a result of ligand diffusion. (c) Pulse signal approaches the channel end leaving behind ligands bound to receptors.}
    \label{smoldyn_MC}
\end{figure}

The simulation setup consisted of a straight microfluidic channel with a rectangular cross-section, as shown in Fig. \ref{smoldyn_MC}. The receptor molecules were immobilized at the channel bed, representing the 2d MC receiver. An input rectangular pulse signal composed of ligands was introduced at the inlet of the channel as shown in Fig. \ref{smoldyn_MC}(a). These ligands propagated towards the receptors through convection and diffusion as depicted in Fig. \ref{smoldyn_MC}(b). A fraction of the ligands randomly bound to the receptor molecules for varying durations, depending on their kinetic interaction rates. Subsequently, they unbound and continued their propagation until they exited the channel at the outlet, as demonstrated in Fig. \ref{smoldyn_MC}(c). 
To validate the model in both time and frequency domains, we evaluated the transfer function of the propagation channel, the transfer function of the LR binding process, and the end-to-end frequency-domain model under varying system parameters. This evaluation was conducted using both analytical expressions and simulation results. The particle-based simulation does not incorporate the transfer function of the MC receiver. Therefore, the numerical results for $H_\mathrm{t}(f)$ are solely obtained using analytical expressions. Moreover, we calculated the sampling frequency utilizing a numerical method for varying system parameters.

\begin{center}
\begin{table}[!t]	
\caption{Default Values of Simulation Parameters}
\label{tab:parameters}
\footnotesize	
    \begin{tabular}{m{6cm}|m{2 cm}}
     \hline \hline 
       Admittance for CPE$_\mathrm{g-e}$ $(Q_\mathrm{g-e})$   &    1.6 $\mu$Fs$^{\alpha-1}$/cm$^2$ \\ 
        \hline
Constant angle of impedance for CPE$_\mathrm{g-e}$ $(\alpha_\mathrm{g-e})$ & 0.905\\
          \hline
         Admittance for CPE$_\mathrm{par}$ $(Q_\mathrm{par})$ & 8 nFs$^{\alpha-1}$\\
           \hline
Constant angle of impedance for CPE$_\mathrm{par}$            $(\alpha_\mathrm{par})$&   0.6\\
            \hline
     Admittance for CPE$_\mathrm{l-e}$ $(Q_\mathrm{l-e})$ & 5.4 fFs$^{\alpha-1}$\\
           \hline
     Constant angle of impedance for CPE$_\mathrm{l-e}$            $(\alpha_\mathrm{l-e})$&   1\\
            \hline
            Graphene channel width $(w_\mathrm{g})$ & 1 $\mu$m\\
            \hline
            Graphene channel length $(l_\mathrm{g})$ & 3 $\mu$m\\
            \hline 
            Drain to source voltage $(V_\mathrm{ds})$ & 0.1 V\\
            \hline 
            Mobility of graphene $(\mu_\mathrm{g})$ & 2$\times$10$^3$ cm$^2$/Vs\\
            \hline
             Ionic strength of electrolyte medium $(c_\mathrm{ion})$ & 0.5 mM \\
        \hline
              Relative permitivity of medium  $(\epsilon_\mathrm{M}/\epsilon_0)$ & 80\\
        \hline
              Temperature $(T)$ & 300 K\\
        \hline
        Length of a surface receptor $(r)$ & 2 nm\\
            \hline
            Diffusion coefficient of ligands $(D)$ &  10 $^{-11}$ m$^2$/s\\
            \hline
            Microfluidic channel height $(h_\mathrm{ch})$ & 3  $\mu$m\\
            \hline 
            Microfluidic channel width $(w_\mathrm{ch})$ & 3  $\mu$m\\
            \hline 
              Microfluidic channel length ($l_\mathrm{ch})$ & 200 $\mu$m\\
            \hline 
            Transmitter-receiver distance $(x_\mathrm{r})$ & $100 $ $\mu$m\\
            \hline
            Flow velocity $(u)$ & $2 \times 10^{-3} $ m/s\\
            \hline 
            Binding rate $(k_\mathrm{+})$ & 10$^{-18} $ m$^3$/s\\
            \hline
            Unbinding rate $(k_\mathrm{-})$ & 500  s$^{-1}$\\
            \hline 
             Pulse concentration $(C_\mathrm{m})$ & $3.3 \times 10^{20}$ m$^{-3}$\\
             \hline
            Average number of electrons in a ligand $(N_\mathrm{e^{-}})$ & 3\\
            \hline
            Number of receptors on the sensor surface $(N_\mathrm{r})$ & 500 \\
            \hline 
            Pulse width $(T_\mathrm{p})$ & 0.5 ms\\
            \hline 
            Simulation time step $(\Delta t)$ & 50 $\mu$s\\
            \hline 
            
    \end{tabular}
\end{table}
\end{center}

\subsection{Propagation Channel}

\subsubsection{Effect of Varying Pulse Width}
\label{varying_PW}
\begin{figure*}
     \centering
     \begin{subfigure}[b]{0.46\textwidth}
         \centering
         \includegraphics[width=\textwidth]{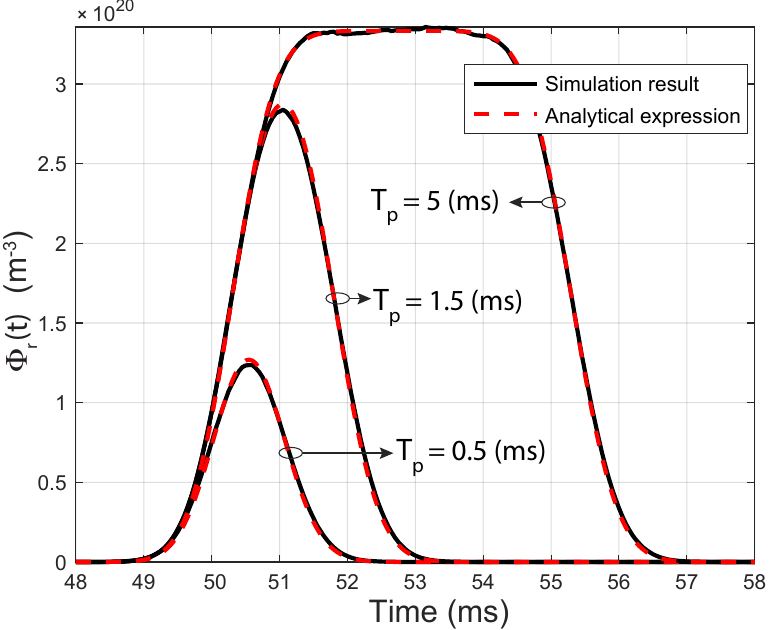}
         \caption{}
         \label{}
     \end{subfigure}
\hfill
     \begin{subfigure}[b]{0.46\textwidth}
         \centering
         \includegraphics[width=\textwidth]{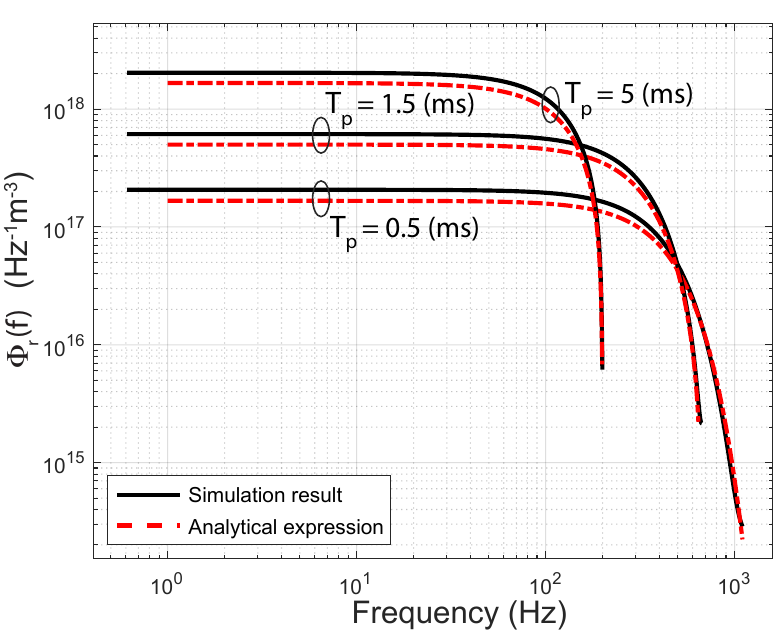}
         \caption{}
         \label{}
     \end{subfigure}
      \caption{(a) The analytical and simulation results of time-domain channel response under varying pulse width. (b) Spectral density of pulse concentrations signal with varying pulse width driven from analytical model and simulation results.}
        \label{fig_PW}
\end{figure*}

The first analysis investigates the impact of varying pulse width, $T_\mathrm{p}$, a critical parameter commonly employed in signal generation and modulation schemes, such as pulse width modulation (PWM) \cite{garrison2018electrically}. The results of this analysis are presented in Fig. \ref{fig_PW}. As expected, an increase in pulse width leads to a higher concentration value in time domain, as shown in Fig. \ref{fig_PW}(a). In the frequency domain (Fig. \ref{fig_PW}(b)), a higher amplitude is observed in the spectral density of the received concentration, $\Phi_\mathrm{r}(f)$, as the pulse width increases. Moreover, the cutoff frequency decreases with the increasing pulse width, which is consistent with the expectations based on Equations \eqref{t_con_out} and \eqref{f_rect}. The analytical model exhibits a high degree of agreement with the simulation results. It is important to note that as pulse width increases, the likelihood of inter-symbol interference also rises, leading to potential challenges in the signal recovery process.
\vspace{5mm}

\subsubsection{Effect of Varying Diffusion Coefficient}
\begin{figure*}[t]
     \centering
     \begin{subfigure}[b]{0.46\textwidth}
         \centering
         \includegraphics[width=\textwidth]{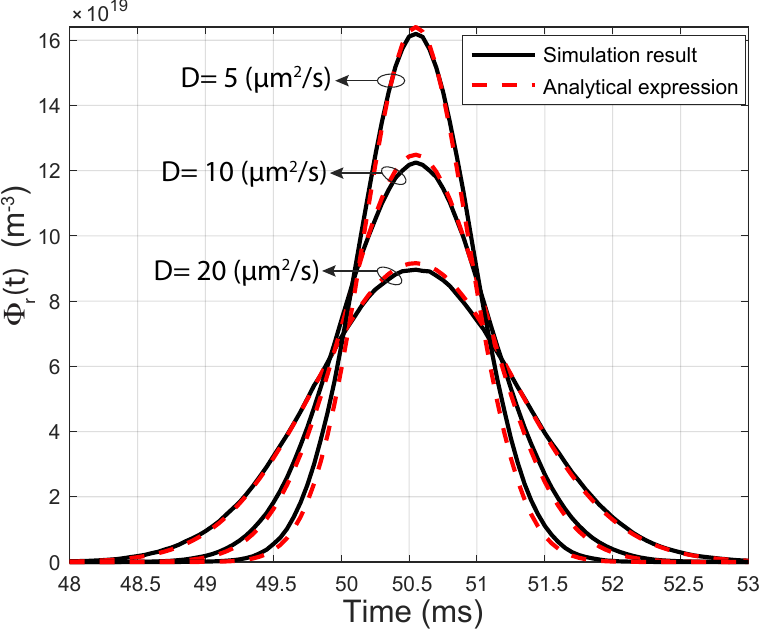}
         \caption{}
         \label{}
     \end{subfigure}
\hfill
     \begin{subfigure}[b]{0.46\textwidth}
         \centering
         \includegraphics[width=\textwidth]{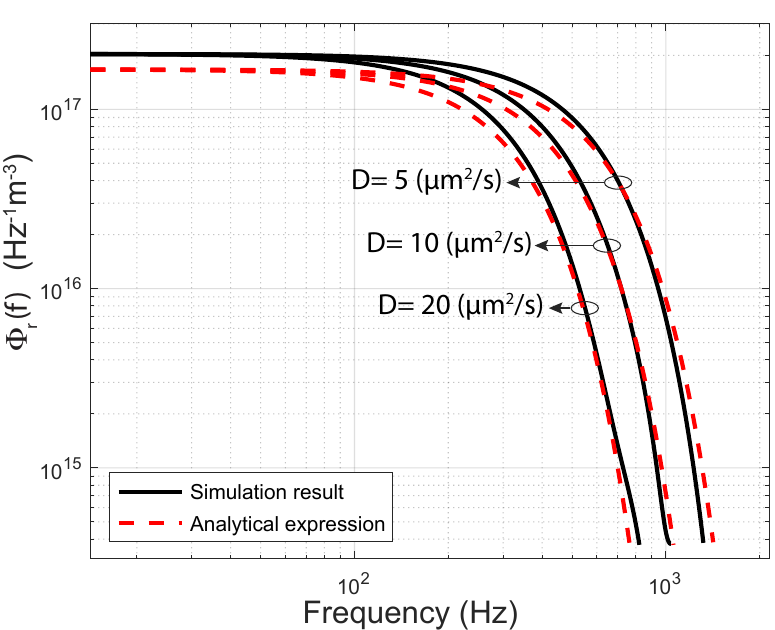}
         \caption{}
         \label{}
     \end{subfigure}

        \caption{(a) Concentration of pulse signals with varying diffusion coefficient in time domain. (b) Spectral density of pulse concentration in a varying diffusion coefficient setting.}
        \label{fig_D}
\end{figure*}

\vspace{5mm}
We also analyze the impact of varying diffusion coefficient of ligands, $D$, on the response of the MC system. The diffusion coefficient is a fundamental parameter in MC systems, as it determines the rate at which molecules disperse through the medium. Molecules with a higher diffusion coefficient disperse more, resulting in a broader received signal width. Additionally, the peak received concentration decreases due to the higher dispersion, as shown in Fig. \ref{fig_D}(a). On the other hand, in the frequency domain, increasing the diffusion coefficient is reflected in a slight decrease in the cutoff frequency of the spectral density of the received concentration, $\Phi_\mathrm{r}(f)$, for both analytical and simulation results, as demonstrated in Fig. \ref{fig_D}(b).
\subsection{Ligand-Receptor Binding Process}
\subsubsection{Effect of Varying Binding Rate} 
\begin{figure*}
     \centering
     \begin{subfigure}[b]{0.46\textwidth}
         \centering
         \includegraphics[width=\textwidth]{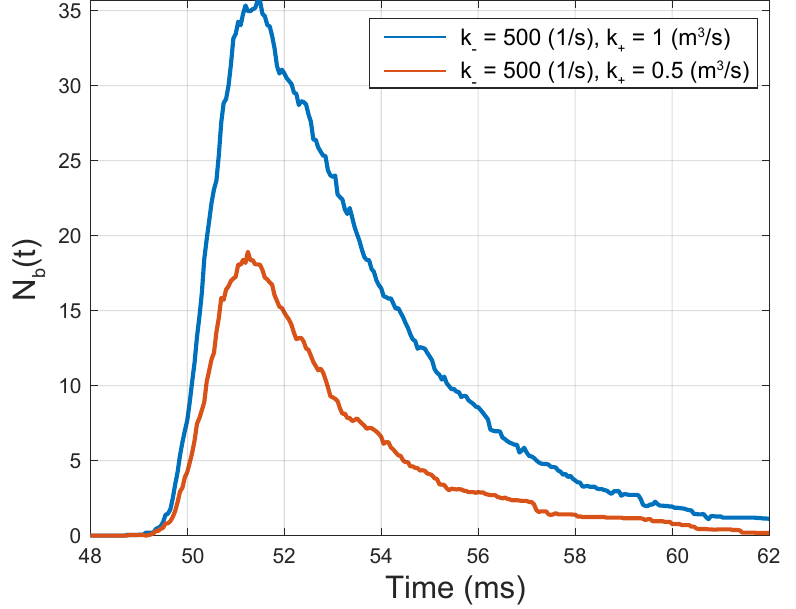}
         \caption{}
         \label{}
     \end{subfigure}
\hfill
     \begin{subfigure}[b]{0.46\textwidth}
         \centering
         \includegraphics[width=\textwidth]{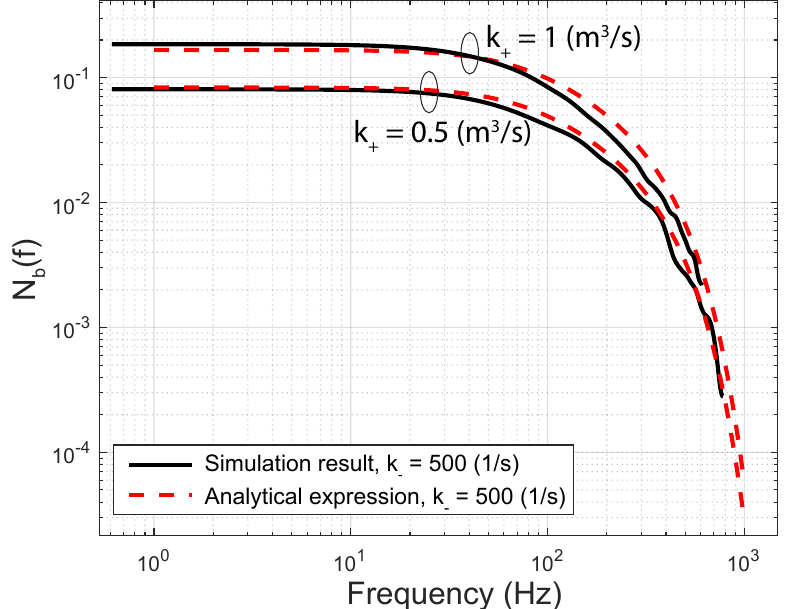}
         \caption{}
         \label{}
     \end{subfigure}

        \caption{(a) Number of bound receptors for varying binding rate in time-domain. (b) Spectral density of number of bound receptors for varying binding rate}
        \label{fig_Bind}
\end{figure*}

We investigate the effect of varying binding rates, $k_\mathrm{+}$, on the time-varying number of bound receptors in both time and frequency domains. Fig. \ref{fig_Bind}(a) demonstrates that an increase in the binding rate directly corresponds to an increased number of bound receptors, $N_\mathrm{b}(t)$, as molecules with higher binding rate exhibit a higher propensity to bind to the receptors when they are in close proximity of each other. Similarly, as expected from Equation \eqref{eq:number_bound}, the frequency domain analysis reveals a higher amplitude in the spectral density of the number of bound receptors, $N_\mathrm{b}(f)$, when binding rates are increased, as shown in Fig. \ref{fig_Bind}(b).

\subsubsection{Effect of Varying Unbinding Rate} 
\begin{figure*}
     \centering
     \begin{subfigure}[b]{0.46\textwidth}
         \centering
         \includegraphics[width=\textwidth]{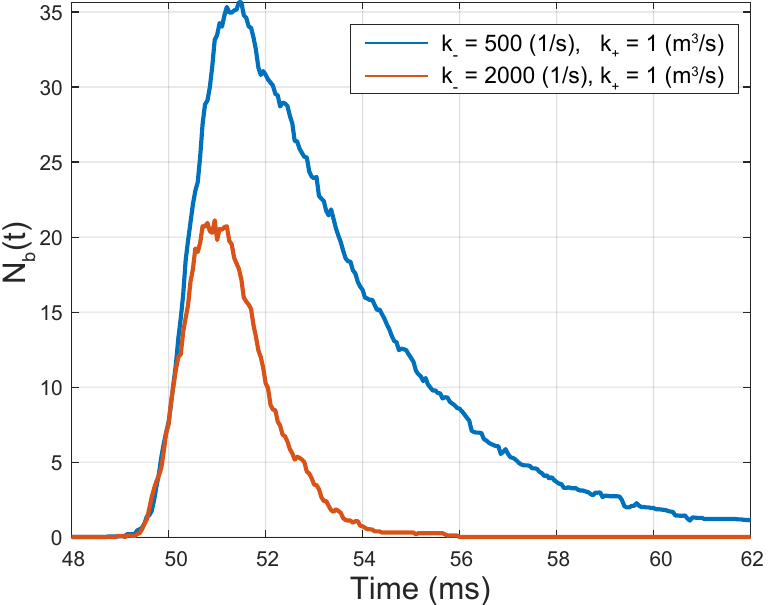}
         \caption{}
         \label{}
     \end{subfigure}
\hfill
     \begin{subfigure}[b]{0.46\textwidth}
         \centering
         \includegraphics[width=\textwidth]{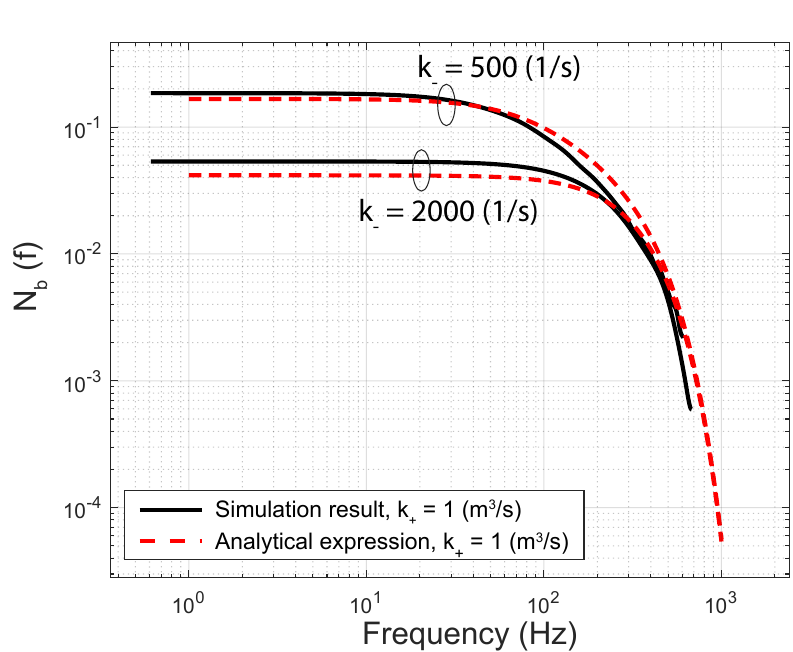}
         \caption{}
         \label{}
     \end{subfigure}

        \caption{(a) Number of bound receptors for varying unbinding rate in time-domain. (b) Spectral density of the number of bound receptors for varying unbinding rate.}
        \label{fig_unbind}
\end{figure*}

We also investigate the impact of varying unbinding rates, $k_\mathrm{-}$, on the number of bound receptors. Contrary to the effect of binding rates, increasing the unbinding rate leads to a decrease in the number of bound receptors in the time-domain, $N_\mathrm{b}(t)$, as shown in \ref{fig_unbind}(a). Molecules with higher unbinding rate have shorter bound state durations. In the frequency domain, as shown in Fig. \ref{fig_unbind}(b), the unbinding rate exhibits an inverse relationship with the spectral density, as described by \eqref{eq:number_bound}. Consequently, a higher unbinding rate results in a lower amplitude in the spectral density of bound receptors, $N_\mathrm{b}(f)$, a finding supported by both simulation and analytical results.  
\vspace{5mm}

\subsection{End-to-End Model}
\subsubsection{Effect of Varying Pulse Width}
\begin{figure*}
     \centering
     \begin{subfigure}[b]{0.46\textwidth}
         \centering
         \includegraphics[width=\textwidth]{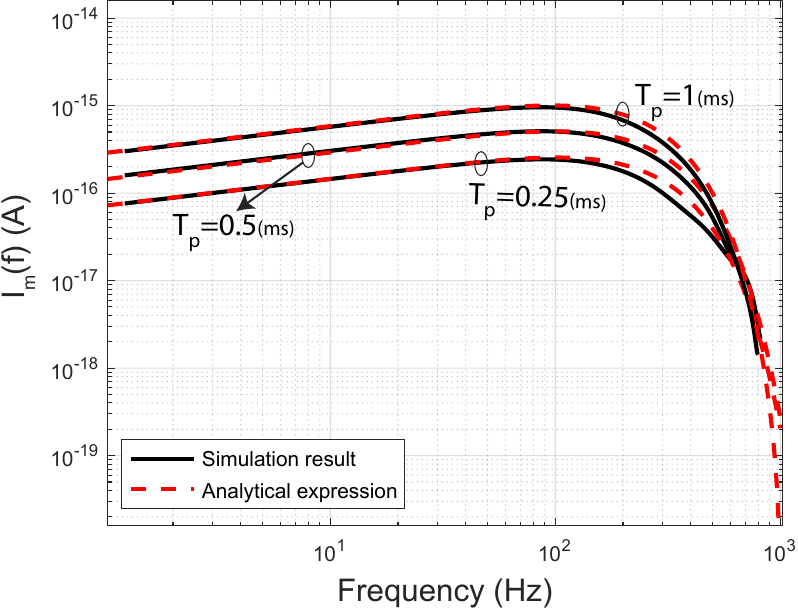}
         \caption{}
         \label{}
     \end{subfigure}
\hfill
     \begin{subfigure}[b]{0.46\textwidth}
         \centering
         \includegraphics[width=\textwidth]{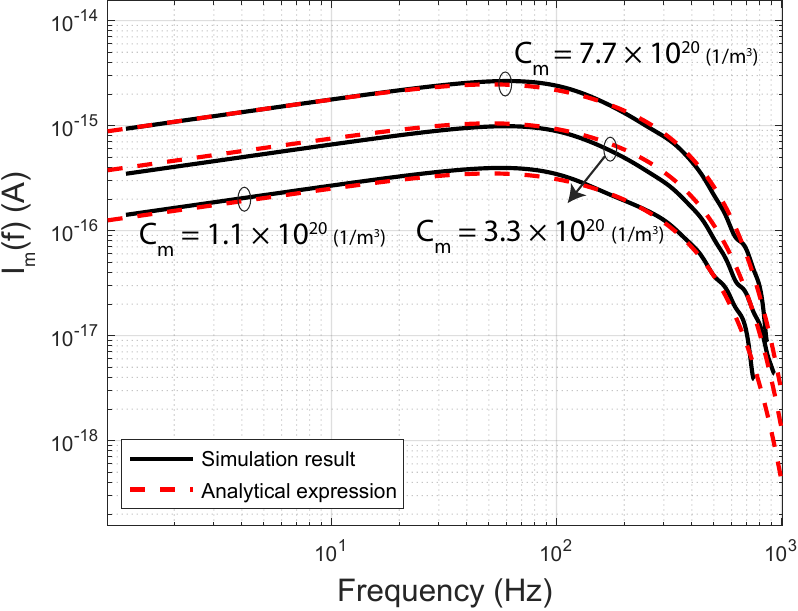}
         \caption{}
         \label{}
     \end{subfigure}

        \caption{Spectral density of output current (end-to-end model) from analytical expression and simulation results (a) for varying pulse width $T\mathrm{p}$, (b) for varying amplitude of input concentration pulse $C_\mathrm{m}$.}
        \label{fig_end}
\end{figure*}

To evaluate the end-to-end model's accuracy and investigate the impact of varying pulse widths, $T_\mathrm{p}$, we analyze the spectral density of output current, $I_\mathrm{m}(f)$, for three pulse signals with different pulse widths but identical amplitudes, i.e., concentrations, as shown in Fig. \ref{fig_end}(a). As predicted in Section \ref{varying_PW}, an increase in pulse width corresponds to a higher amplitude in $I_\mathrm{m}(f)$. This phenomenon occurs  because a wider ligand pulse results in a higher concentration of ligands in the vicinity of the receiver's receptors. This, in turn, increases the probability of binding to a receptor before the already bound ones dissociate, resulting in a higher number of observed bound receptors, i.e., amplitude. The analytical expression for the output current spectrum, represented by \eqref{Im} and incorporating the transfer function of the three main processes and the input signal concentration, demonstrates high accuracy when compared to the simulation results.

\subsubsection{Effect of Varying Ligand Concentration}
We also evaluate the impact of varying ligand concentrations, $C_\mathrm{m}$, on the end-to-end model by performing simulations with input concentration pulses with different concentrations but identical pulse widths. By analyzing the spectral density of the resulting output current, $I_\mathrm{m}(f)$, we observe that the amplitude of $I_\mathrm{m}(f)$ increases as concentration increases, as depicted in Fig. \ref{fig_end}(b).  The simulation results are strongly aligned with the analytical results obtained from \eqref{Im}.

\subsection{Sampling Frequency}
\begin{figure*}[t]
     \centering
     \begin{subfigure}[b]{0.3\textwidth}
         \centering
         \includegraphics[width=\textwidth]{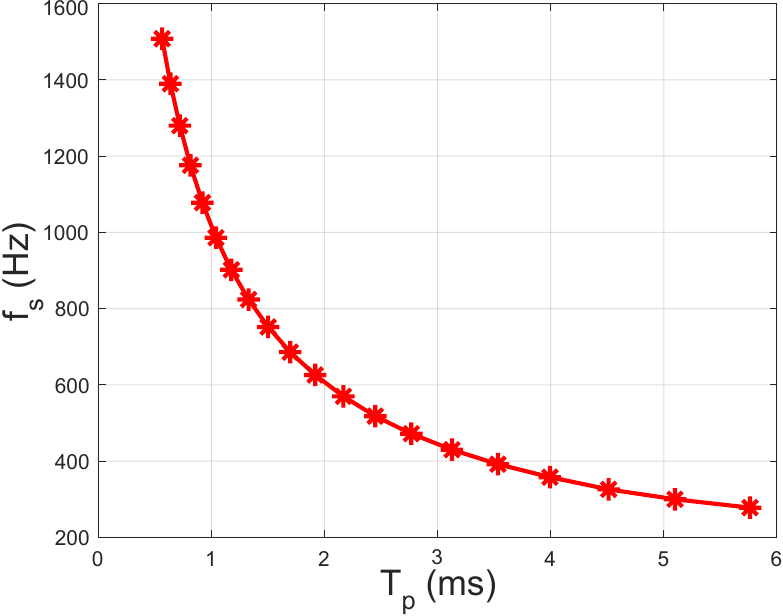}
         \caption{}
         \label{}
     \end{subfigure}
     \hfill
     \begin{subfigure}[b]{0.3\textwidth}
         \centering
         \includegraphics[width=\textwidth]{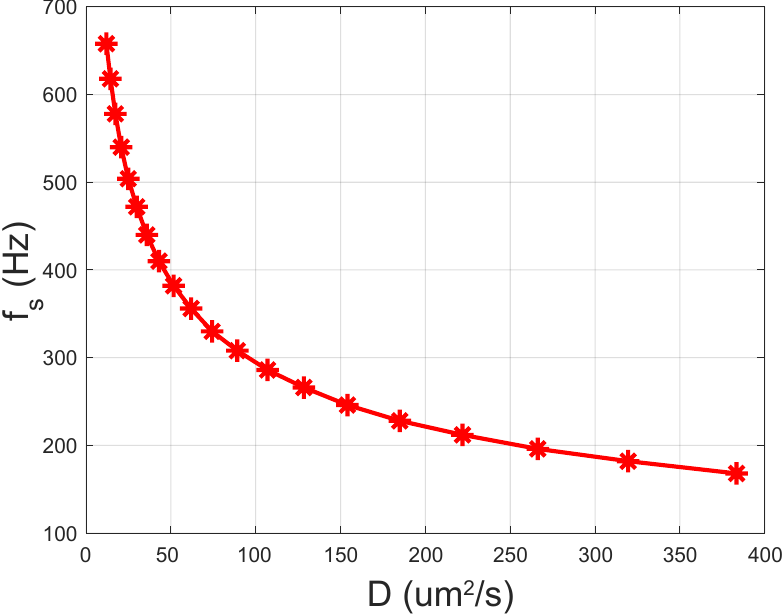}
         \caption{}
         \label{}
     \end{subfigure}
     \hfill
     \begin{subfigure}[b]{0.3\textwidth}
         \centering
         \includegraphics[width=\textwidth]{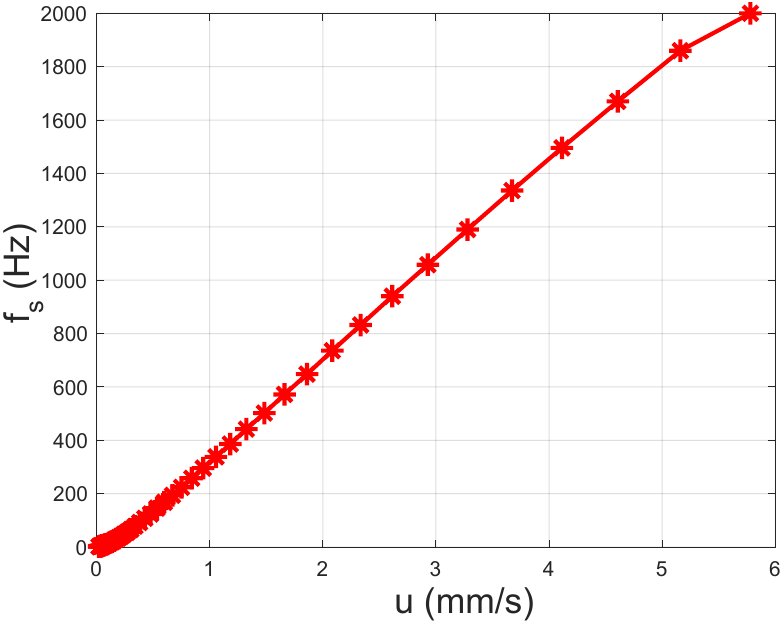}
         \caption{}
         \label{}
     \end{subfigure}
        \caption{Sampling frequency $f_\mathrm{s}$ as a function of (a) pulse width $T_\mathrm{p}$, (b) diffusion coefficient $D$, and (c) flow velocity $u$.}
        \label{sample}
\end{figure*}

 To reconstruct the input concentration signal, $\phi_\mathrm{in}(t)$, from the sampled sequence of the number of bound receptors, it is essential to employ an appropriate sampling frequency. Considering that both the input concentration spectral density and the end-to-end transfer function and consequently the resulting output current spectral density, display a Lorentzian-shaped profile, it is essential to determine the cutoff frequency that contains most of the spectrum energy. The energy of the output current spectral density within a bandwidth ranging from 0 Hz to the cutoff frequency can be quantified as follows \cite{huang2021ffrequency}:
\begin{align}
\int^{f_\mathrm{c}}_{0} |H(f)  \Phi_\mathrm{in}(f)|^2 df =\eta \int^{+ \infty}_{0} |H(f)  \Phi_\mathrm{in}(f)|^2 df, 
\label{cutoff}
\end{align}
where $f_c$ is the cutoff frequency and $\eta$ is the fraction of the total spectrum energy contained within the interval $(0,f_\mathrm{c})$. In this study, we consider $\eta = 0.99$, which indicates that $99\%$ of the spectral power is contained within the specified bandwidth. Once the cutoff frequency has been determined, the sampling frequency can be obtained using the Nyquist–Shannon theorem, which states that in order to achieve a reconstruction that captures all the information, the sampling frequency should be greater than twice the bandwidth:
\begin{align}
    2f_\mathrm{c}\leq f_\mathrm{s} \leq \infty.
    \label{fc}
\end{align}
Fig. \ref{sample} shows the sampling frequency obtained from \eqref{fc}, which is a function of pulse width $T_\mathrm{p}$, diffusion coefficient $D$, and flow velocity $u$. Fig. \ref{sample}(a) demonstrates that increasing the pulse width results in a lower sampling frequency required to reconstruct the original continuous signal. As shown in Fig. \ref{fig_PW}(b), the spectral density of a wider pulse signal has a lower cutoff frequency. Therefore, it is expected that increasing the pulse width would allow a lower sampling frequency. 

Fig. \ref{sample}(b) indicates that the sampling frequency decreases as the diffusion coefficient increases. This can be attributed to the reduction in the cutoff frequency while increasing the diffusion coefficient, as shown in Fig. \ref{fig_D}(b). Therefore, the decrease in sampling frequency aligns with the expectations set by the Nyquist-Shannon theorem. 

Finally, Fig. \ref{sample}(c) shows the impact of increasing flow velocity on the sampling frequency. As the flow velocity increases, the signals traverse the receiver position more quickly, which reduces the time window available for capturing an adequate number of samples from the propagating signal. Consequently, to guarantee the collection of a sufficient number of samples, it is necessary to raise the sampling frequency in response to an increase in flow velocity.

\section{Conclusion} \label{conclusion}
In this study, we introduced a comprehensive end-to-end frequency-domain model for a practical microfluidic MC system with a graphene bioFET-based receiver. The model provides valuable insights into the dispersion and distortion of received signals, and has the potential to inform the design of new frequency-domain MC techniques, such as modulation and detection, matched filters, and interference-free receiver architectures. The end-to-end transfer function, denoted as $H(f)$, incorporates the input-output relationships of three sequential modules: the microfluidic propagation channel, the LR binding process, and the graphene bioFET-based receiver. The accuracy and reliability of the developed model were verified through particle-based spatial stochastic simulations, which demonstrated a high degree of agreement with the analytical expressions.

\bibliographystyle{IEEEtran}

\bibliography{references}

\begin{thebibliography}{10}
\providecommand{\url}[1]{#1}
\csname url@rmstyle\endcsname
\providecommand{\newblock}{\relax}
\providecommand{\bibinfo}[2]{#2}
\providecommand\BIBentrySTDinterwordspacing{\spaceskip=0pt\relax}
\providecommand\BIBentryALTinterwordstretchfactor{4}
\providecommand\BIBentryALTinterwordspacing{\spaceskip=\fontdimen2\font plus
\BIBentryALTinterwordstretchfactor\fontdimen3\font minus
  \fontdimen4\font\relax}
\providecommand\BIBforeignlanguage[2]{{%
\expandafter\ifx\csname l@#1\endcsname\relax
\typeout{** WARNING: IEEEtran.bst: No hyphenation pattern has been}%
\typeout{** loaded for the language `#1'. Using the pattern for}%
\typeout{** the default language instead.}%
\else
\language=\csname l@#1\endcsname
\fi
#2}}

\bibitem{akan2016fundamentals}
O.~B. Akan, H.~Ramezani, T.~Khan, N.~A. Abbasi, and M.~Kuscu, ``Fundamentals of
  molecular information and communication science,'' \emph{Proceedings of the
  IEEE}, vol. 105, no.~2, pp. 306--318, 2016.

\bibitem{akyildiz2015internet}
I.~F. Akyildiz, M.~Pierobon, S.~Balasubramaniam, and Y.~Koucheryavy, ``The
  internet of bio-nano things,'' \emph{IEEE Communications Magazine}, vol.~53,
  no.~3, pp. 32--40, 2015.

\bibitem{akyildiz2020panacea}
I.~F. Akyildiz, M.~Ghovanloo, U.~Guler, T.~Ozkaya-Ahmadov, A.~F. Sarioglu, and
  B.~D. Unluturk, ``Panacea: An internet of bio-nanothings application for
  early detection and mitigation of infectious diseases,'' \emph{IEEE Access},
  vol.~8, pp. 140\,512--140\,523, 2020.

\bibitem{kuscu2019transmitter}
M.~Kuscu, E.~Dinc, B.~A. Bilgin, H.~Ramezani, and O.~B. Akan, ``Transmitter and
  receiver architectures for molecular communications: A survey on physical
  design with modulation, coding, and detection techniques,'' \emph{Proceedings
  of the IEEE}, vol. 107, no.~7, pp. 1302--1341, 2019.

\bibitem{jamali2019channel}
V.~Jamali, A.~Ahmadzadeh, W.~Wicke, A.~Noel, and R.~Schober, ``Channel modeling
  for diffusive molecular communication—a tutorial review,''
  \emph{Proceedings of the IEEE}, vol. 107, no.~7, pp. 1256--1301, 2019.

\bibitem{zadeh2023microfluidic}
M.~K. Zadeh, I.~M. Bolhassan, and M.~Kuscu, ``Microfluidic pulse shaping
  methods for molecular communications,'' \emph{Nano Communication Networks},
  p. 100453, 2023.

\bibitem{civas2023graphene}
\BIBentryALTinterwordspacing
M.~Civas, M.~Kuscu, O.~Cetinkaya, B.~E. Ortlek, and O.~B. Akan, ``Graphene and
  related materials for the internet of bio-nano things,'' \emph{in print, APL
  Materials}, 2023. [Online]. Available: \url{https://arxiv.org/abs/2304.03824}
\BIBentrySTDinterwordspacing

\bibitem{kuscu2021fabrication}
M.~Kuscu, H.~Ramezani, E.~Dinc, S.~Akhavan, and O.~B. Akan, ``Fabrication and
  microfluidic analysis of graphene-based molecular communication receiver for
  internet of nano things (iont),'' \emph{Scientific Reports}, vol.~11, no.~1,
  pp. 1--20, 2021.

\bibitem{pierobon2010physical}
M.~Pierobon and I.~F. Akyildiz, ``A physical end-to-end model for molecular
  communication in nanonetworks,'' \emph{IEEE Journal on Selected Areas in
  Communications}, vol.~28, no.~4, pp. 602--611, 2010.

\bibitem{huang2021frequency}
Y.~Huang, F.~Ji, Z.~Wei, M.~Wen, X.~Chen, Y.~Tang, and W.~Guo, ``Frequency
  domain analysis and equalization for molecular communication,'' \emph{IEEE
  Transactions on Signal Processing}, vol.~69, pp. 1952--1967, 2021.

\bibitem{wang2014transmit}
S.~Wang, W.~Guo, and M.~D. McDonnell, ``Transmit pulse shaping for molecular
  communications,'' in \emph{2014 IEEE Conference on Computer Communications
  Workshops (INFOCOM WKSHPS)}.\hskip 1em plus 0.5em minus 0.4em\relax IEEE,
  2014, pp. 209--210.

\bibitem{civas2023frequency}
\BIBentryALTinterwordspacing
M.~Civas, A.~Abdali, M.~Kuscu, and O.~B. Akan, ``Frequency-domain detection for
  molecular communications,'' \emph{Proceedings of IEEE International Confrence
  on Communication (ICC)}, 2023. [Online]. Available:
  \url{https://arxiv.org/abs/2301.01049}
\BIBentrySTDinterwordspacing

\bibitem{smoldyn2022}
S.~S. Andrews, ``Smoldyn: particle-based simulation with rule-based modeling,
  improved molecular interaction and a library interface,''
  \emph{Bioinformatics}, vol.~33, no.~5, pp. 710--717, 2017.

\bibitem{kuscu2018modeling}
M.~Kuscu and O.~B. Akan, ``Modeling convection-diffusion-reaction systems for
  microfluidic molecular communications with surface-based receivers in
  internet of bio-nano things,'' \emph{PloS One}, vol.~13, no.~2, p. e0192202,
  2018.

\bibitem{bicen2013system}
A.~O. Bicen and I.~F. Akyildiz, ``System-theoretic analysis and least-squares
  design of microfluidic channels for flow-induced molecular communication,''
  \emph{IEEE Transactions on Signal Processing}, vol.~61, no.~20, pp.
  5000--5013, 2013.

\bibitem{kuscu2019channel}
M.~Kuscu and O.~B. Akan, ``Channel sensing in molecular communications with
  single type of ligand receptors,'' \emph{IEEE Transactions on
  Communications}, vol.~67, no.~10, pp. 6868--6884, 2019.

\bibitem{lauffenburger1996receptors}
D.~A. Lauffenburger and J.~Linderman, \emph{Receptors: models for binding,
  trafficking, and signaling}.\hskip 1em plus 0.5em minus 0.4em\relax Oxford
  University Press, 1996.

\bibitem{shahmohammadian2013modelling}
H.~ShahMohammadian, G.~G. Messier, and S.~Magierowski, ``Modelling the
  reception process in diffusion-based molecular communication channels,'' in
  \emph{2013 IEEE International Conference on Communications Workshops
  (ICC)}.\hskip 1em plus 0.5em minus 0.4em\relax IEEE, 2013, pp. 782--786.

\bibitem{garcia2020distortion}
R.~Garcia-Cortadella, E.~Masvidal-Codina, J.~M. De~la Cruz, N.~Sch{\"a}fer,
  G.~Schwesig, C.~Jeschke, J.~Martinez-Aguilar, M.~V. Sanchez-Vives, R.~Villa,
  X.~Illa, \emph{et~al.}, ``Distortion-free sensing of neural activity using
  graphene transistors,'' \emph{Small}, vol.~16, no.~16, p. 1906640, 2020.

\bibitem{stojek2010electrical}
Z.~Stojek, ``The electrical double layer and its structure,''
  \emph{Electroanalytical methods: Guide to experiments and applications}, pp.
  3--9, 2010.

\bibitem{khademi2020structure}
M.~Khademi and D.~P. Barz, ``Structure of the electrical double layer
  revisited: Electrode capacitance in aqueous solutions,'' \emph{Langmuir},
  vol.~36, no.~16, pp. 4250--4260, 2020.

\bibitem{sun2019unique}
J.~Sun and Y.~Liu, ``Unique constant phase element behavior of the
  electrolyte--graphene interface,'' \emph{Nanomaterials}, vol.~9, no.~7, p.
  923, 2019.

\bibitem{barsoukov2005impedance}
E.~Barsoukov and J.~R. Macdonald, ``Impedance spectroscopy theory, experiment,
  and,'' \emph{Applications, 2nd ed.(Hoboken, NJ: John Wiley \&Sons, Inc.,
  2005)}, 2005.

\bibitem{hsu2001concerning}
C.~Hsu and F.~Mansfeld, ``Concerning the conversion of the constant phase
  element parameter y0 into a capacitance,'' \emph{Corrosion}, vol.~57, no.~09,
  2001.

\bibitem{xu2017real}
S.~Xu, J.~Zhan, B.~Man, S.~Jiang, W.~Yue, S.~Gao, C.~Guo, H.~Liu, Z.~Li,
  J.~Wang, \emph{et~al.}, ``Real-time reliable determination of binding
  kinetics of dna hybridization using a multi-channel graphene biosensor,''
  \emph{Nature Communications}, vol.~8, no.~1, p. 14902, 2017.

\bibitem{xu2011top}
H.~Xu, Z.~Zhang, H.~Xu, Z.~Wang, S.~Wang, and L.-M. Peng, ``Top-gated graphene
  field-effect transistors with high normalized transconductance and designable
  dirac point voltage,'' \emph{ACS Nano}, vol.~5, no.~6, pp. 5031--5037, 2011.

\bibitem{kuscu2016modeling}
M.~Kuscu and O.~B. Akan, ``Modeling and analysis of sinw fet-based molecular
  communication receiver,'' \emph{IEEE Transactions on Communications},
  vol.~64, no.~9, pp. 3708--3721, 2016.

\bibitem{rajan2013performance}
N.~K. Rajan, X.~Duan, and M.~A. Reed, ``Performance limitations for
  nanowire/nanoribbon biosensors,'' \emph{Wiley Interdisciplinary Reviews:
  Nanomedicine and Nanobiotechnology}, vol.~5, no.~6, pp. 629--645, 2013.

\bibitem{kuscu2016physical}
M.~Kuscu and O.~B. Akan, ``On the physical design of molecular communication
  receiver based on nanoscale biosensors,'' \emph{IEEE Sensors Journal},
  vol.~16, no.~8, pp. 2228--2243, 2016.

\bibitem{pierobon2011noise}
M.~Pierobon and I.~F. Akyildiz, ``Noise analysis in ligand-binding reception
  for molecular communication in nanonetworks,'' \emph{IEEE Transactions on
  Signal Processing}, vol.~59, no.~9, pp. 4168--4182, 2011.

\bibitem{garrison2018electrically}
J.~Garrison, Z.~Li, B.~Palanisamy, L.~Wang, and E.~Seker, ``An
  electrically-controlled programmable microfluidic concentration waveform
  generator,'' \emph{Journal of biological engineering}, vol.~12, no.~1, pp.
  1--10, 2018.

\bibitem{huang2021ffrequency}
Y.~Huang, F.~Ji, M.~Wen, Y.~Tang, X.~Chen, and W.~Guo, ``A frequency domain
  view on diffusion-based molecular communication channels,'' in \emph{ICC
  2021-IEEE International Conference on Communications}.\hskip 1em plus 0.5em
  minus 0.4em\relax IEEE, 2021, pp. 1--6.

\end{thebibliography}




\end{document}